\documentclass{article}
\usepackage{emulateapj}

\newcounter{myfigs}

\begin{document}

\title{A Spin Modulated Telescope to Make Two Dimensional CMB Maps}

\author{J. Staren, P. Meinhold, J. Childers, M. Lim\altaffilmark{1}, A. Levy, P. Lubin}
\affil{Physics Department, University of California at Santa Barbara, Santa Barbara, CA  93106-2530}

\author{M. Seiffert, T. Gaier}
\affil{Jet Propulsion Laboratories}

\author{N. Figueiredo}
\affil{Escola Federal de Engenharia de Itajub\'{a}, 37500-000, Itajub\'{a}, MG, Brazil}

\author{T. Villela, C.A. Wuensche}
\affil{Divis\~{a}o de Astrof{\'\i}sica, Instituto National de Pesquisas Espaciais, 12225-010, S\~{a}o Jos\'{e} dos Campos, SP, Brazil}

\author{M. Tegmark and A. de Oliveira-Costa}
\affil{Department of Physics, University of Pennsylvania, Philadelphia, PA 19104}

\altaffiltext{1}{now at Brightmail, Inc
301 Howard Street, Suite 1800
San Francisco, CA 94105}

\begin{abstract}
We describe the HEMT Advanced Cosmic Microwave Explorer (HACME), a balloon borne experiment designed to measure sub-degree scale Cosmic Microwave Background anisotropy over hundreds of square degrees, using a unique two dimensional scanning strategy.  A spinning flat mirror that is canted relative to its spin axis modulates the direction of beam response in a nearly elliptical path on the sky. The experiment was successfully flown in February of 1996, achieving near laboratory performance for several hours at float altitude.  A map free of instrumental systematic effects is produced for a 3.5 hour observation of 630 square degrees, resulting in a flat band power upper limit of $\langle \ell(\ell+1)C_\ell/(2\pi)\rangle^{0.5} < 77~\mu$K at $\ell = 38^{+25}_{-20}$ (95\% confidence).  The experiment design, flight operations and data, including atmospheric effects and noise performance, are discussed.
\end{abstract}

\keywords{cosmic microwave background -- cosmology:  observations -- instrumentation: detectors}

\section{Introduction}
The Cosmic Microwave Background (CMB) is a unique
relic of the primordial Universe and the study of its properties promises
to advance our understanding of the formation of structure in the universe.
Detection of CMB anisotropy helps to discriminate among
potential models with different cosmological parameters
(\cite{Bon97}; \cite{Jun96}; \cite{HuS97}; \cite{Zal97}).
Much of the discriminating power of CMB measurements comes from angular scales below 1 degree, or Legendre multipoles, $l$, greater than 100 in the angular power spectrum.

The effort to understand the power spectrum of the CMB for large $l$ has continued since the discovery of anisotropy in the CMB (\cite{Smo92,Als92,Sch93,Rea89,Gun95,Lim96,Che97,Net97}).  Many previous experiments minimized errors for the best sensitivity at a single angular scale on a small region of sky.  An ideal experiment will have sufficient sky coverage with enough sensitivity over a broad angular range to measure the power spectrum at several angular scales.  Broadly speaking, the optimal strategy is to aim for a signal to noise of order unity per pixel with the largest and most symmetric sky coverage possible, for a given resolution and instrument sensitivity (\cite{Kno97,Teg97A,Teg97B}).  Instrumentation has evolved so that sensitive large area maps can be obtained with sub-degree resolution over a range of microwave frequencies (\cite{Cob99,dOC98,Net97}).  With sensitivity from the beamsize to the large angle of the spinning chop, a multi-pixel spin-modulated telescope easily covers enough sky area to make precision measurements of the angular power spectrum.  In principle, future experiments will enable cosmological parameter extraction from such power spectrum measurements. 

The HACME experiment is the precursor to several spin modulated experiments that will accurately measure the angular power spectrum using cryogenic HEMT (High Electon-Mobility Transistor) amplifier receivers flown on balloon platforms to minimize atmospheric contamination and enable large angle chopping.  HACME has a single Q-band receiver fed by the ACME telescope (\cite{Mei92}) and modulated by a 1.2~m rotating canted flat mirror.  The single horn and conventional duration balloon flight allows an upper limit on CMB anisotropy and upcoming instruments will have more horns and longer flights to make power spectrum estimates.  The novel optical spinning technique, described in detail in \S2.1, allows rapid two dimensional modulation of the instrument to reduce the effects of 1/f noise in the instrument while tying together widely separated regions of the sky in a short time.  Changes in optics, computers and data acquisition made for HACME are also described in \S2.  In \S3, the observations made and the data collected are discussed.

\section{INSTRUMENT}

\subsection{Optics and Scan Strategy}

HACME's optical system (Figure \ref{f_gond}) consists of a fixed ellipsoidal mirror, the original ACME 1~m parabolic primary mirror, a new 1.2~m spinning flat mirror and a baffle surrounding the flat mirror between $2^\circ$ to $5^\circ$ from the optical axes to intercept direct spillover paths from the Earth to the edge of the primary mirror.  The spinning flat mirror, canted $2.5^\circ$ relative to its spin axis, modulates the telescope beam in a nearly elliptical path on the sky so we become sensitive to anisotropy on scales from the beamsize, $0.8^\circ$ FWHM, to the size of the major axis of the ellipse, $10.0^\circ$.  The polar angle, $\theta_r$, and an azimuthal angle about the center of the path, $\phi_r$, are 
\begin{equation}
\cos(\theta_r) = \cos(2\theta_c)~+~2\sin^2(\theta_I)\sin^2(\theta_c)
\sin^2(\phi_f)\label{thetar}\\
\end{equation}
and 
\begin{flushleft}$
\tan(\phi_r) = \nonumber 
$\end{flushleft}
\begin{equation}
\hfil \sin(\phi_f) 
\bigg[{\cos(\theta_c) \cos(\theta_I) - \sin(\theta_c) \sin(\theta_I) \cos(\phi_f)
\over \cos(\theta_c) \cos(\phi_f) + {1\over2} \sin(2\theta_I) \sin(\theta_c) \sin^2(\phi_f)}\bigg],\label{phir}
\end{equation}
where $\theta_c = 2.5^\circ$ is the cant, $\theta_I = 31.9^\circ$ is the angle between the mirror spin axis and the main optical axis and $\phi_f$ is the phase of the flat mirror spin.  
Performing an azimuth scan adds another level of modulation, building up a two dimensional image of the CMB.  At the start of the azimuth scan, the leftmost part of the beam path is aligned with a fiducial star.  The gondola scans $20^\circ$ to the right and returns at a rate of 0.17 degrees per second.  This process continues as we track the fiducial star across the sky.  This observing strategy is shown on a map in Figure \ref{f_map}.  Data taken in this manner produce large maps with nearly diagonal noise covariance matrices (\cite{Wri96}, \cite{Teg97B}).  The resulting region of sky covered is almost $30^\circ$ wide and between $10^\circ$ and $30^\circ$ high for a typical three to five hour observation.

{
\refstepcounter{myfigs}
\label{f_gond}
	\centerline{{\vbox{\epsfxsize=9.5cm\epsfbox{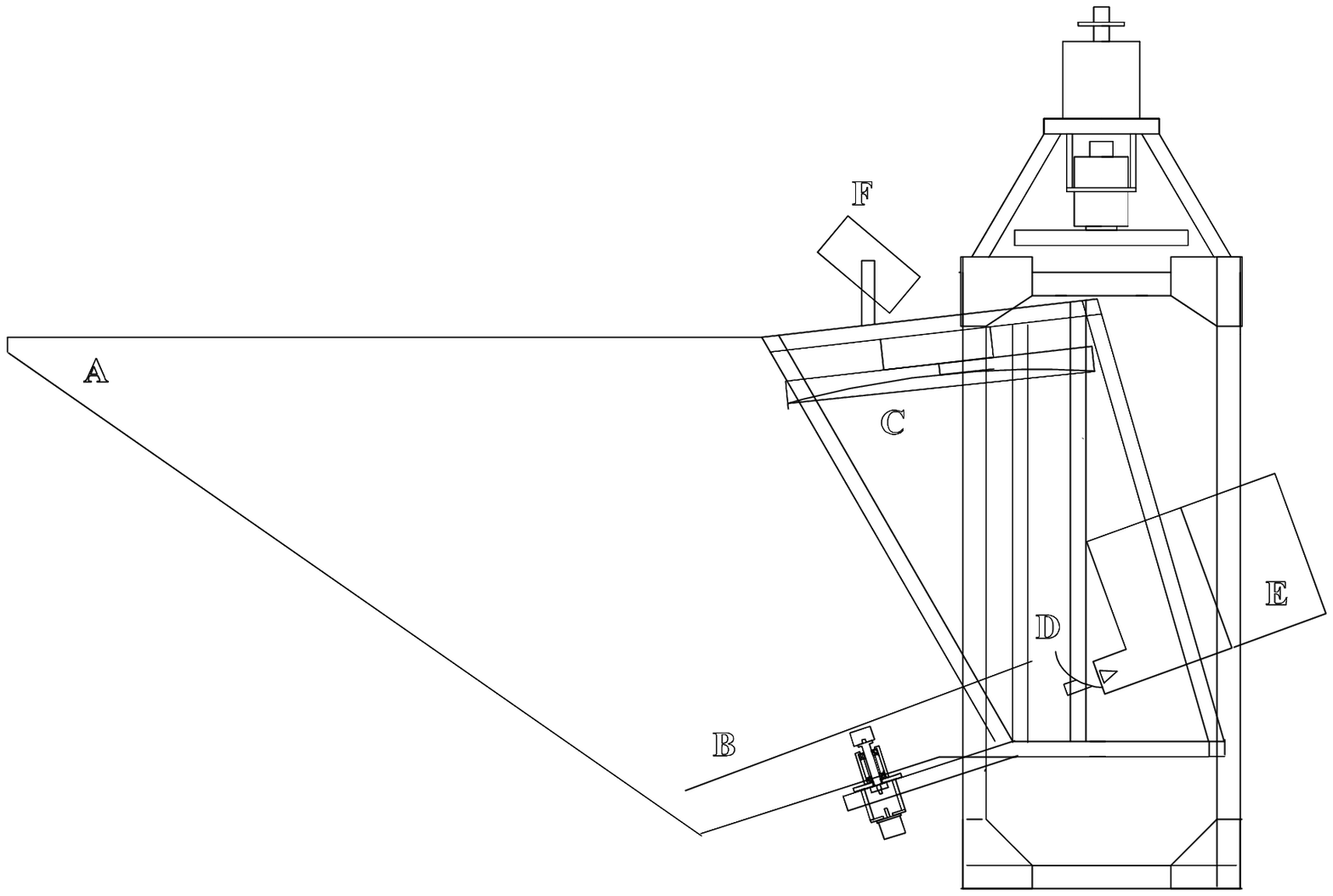}}}}

{\footnotesize{FIG.~\ref{f_gond}.  Schematic Drawing of the Major Optical Elements of HACME:  A. Baffle; B.  Canted Spinning Flat Mirror; C.  1~m Parabolic Mirror; D.  Ellipsoidal Mirror; E.  Dewar; F.  CCD Camera.}}
\bigskip
}

Telescope response is measured on an antenna range in Santa Barbara with a Gunn oscillator source.  The best fit Gaussian to the main lobe along the azimuthal and elevation axes at 40~GHz are $0.84\pm0.02^\circ$ and $0.75\pm0.02^\circ$ FWHM respectively.  The beam shape is independent of the orientation of the flat mirror to within the measurement errors and varies as the inverse of the frequency of the subbands described below.  Observations of the Moon during flight confirm our ground tests of the beam size.  We measured the illumination taper of the primary to be below -25~dB around the edge and calculated the taper around the rotating flat mirror to be always less than -39~dB. Sidelobe measurements made with the main baffle in place demonstrate response below -60~dB for angles greater than $20^\circ$ from boresight along the elevation axis.  

\subsection{Receiver and Data Acquisition}

The HACME Q-band receiver is the same as that of SP94 (\cite{Gun95}), with a five-stage InP-GaAs HEMT multiplexed into three 2.3~GHz bandwidth subbands centered at 39.15, 41.45 and 43.75~GHz, but is used as a total-power radiometer for this instrument.  The output diodes for each subband connect to a low noise pre-amplifier, an A/C coupling stage ($1/RC\approx7$~sec) and a linear voltage to frequency (V/F) converter.  The output of the V/F is counted synchronously with the spin of the flat mirror, 125 triggers per rotation.  An indexed optical encoder attached to the spinning mirror shaft provides the triggers that divide each mirror spin into 125 ``sectors".  The mirror spins at $2.500~\pm~0.004$ Hz, resulting in an ideal integration of the diode outputs at 3.2~msec.  Pulses from a 16~MHz clock are counted for normalization.  Pointing and receiver data are stored on board the payload and a compressed form of the data is sent to the ground.  Analysis is performed on the data recovered from the hard disk after landing. 

\subsection{Calibration}
The telescope is calibrated using two methods.  First, the absolute calibration, including atmospheric attenuation and telescope loss, is determined by observing the Moon.  A model of lunar emission by S. Keihm (JPL) is compared to two-dimensional maps made from the data.  Secondly, an ambient temperature ``Eccosorb" target can intercept the beam between the primary and secondary mirrors to provide relative calibration during the flight.  The latter calibrations are performed about every hour during flight and determine the relative calibration to $\pm3\%$.  The following results are calibrated in antenna temperature units.  Conversion of the data to CMB thermodynamic temperature requires multiplication by 1.04, 1.05 and 1.05 for the 39, 41 and 43~GHz subbands.  

\section{Data}

\subsection{Observations}

This experiment flew for the first time on 1996 February 11 from Ft Sumner NM, obtaining 15 hours at a float altitude of 35.4~km.  Operations proceeded smoothly, with detectors and all servo systems reaching expected performance for the entire flight.  Of the 12 nighttime hours at float, 8 hours were spent mapping three extended regions near the stars Alpha Leonis, Gamma Eridani and Gamma Ursae Minoris.  These regions were chosen for low foreground emission and with systematic tests in mind.  All regions observed are between $40^\circ$ and $50^\circ$ from the galactic plane.  When observed, $\alpha$Leo was rising in the east, $\gamma$Eri was setting in the southwest and $\gamma$UMi was transiting in the north.  The remainder of the time was spent on calibrations, main lobe mapping using the Moon, systematic tests such as operating without the servo systems and RF transmitters.  A second flight was attempted on 1996 June 1 but was terminated by a balloon failure before reaching float altitude.

{
\refstepcounter{myfigs}
\label{f_map}
	\centerline{{\vbox{\epsfxsize=9.5cm\epsfbox{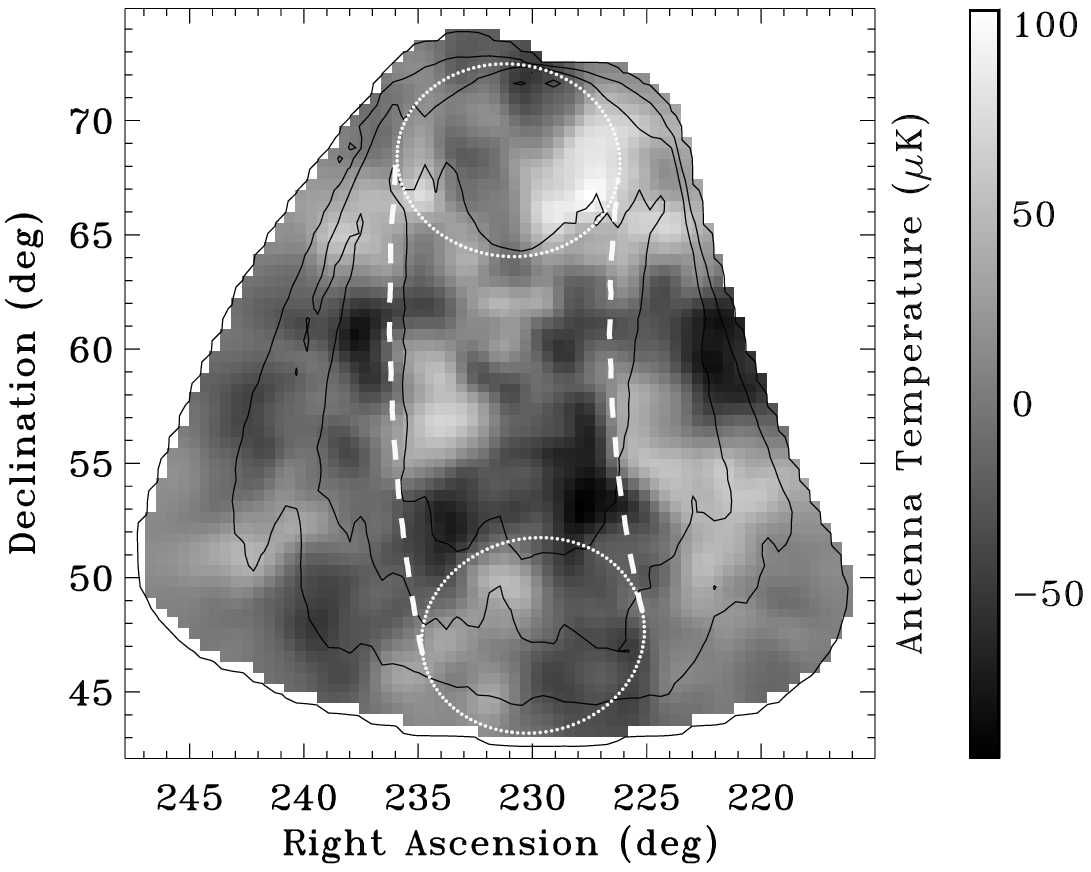}}}}
{\footnotesize{FIG.~\ref{f_map}.  Gamma Ursae Minoris map.  The map is made from the 43~GHz subband data.  There are over 3000 gnomonic 28' square pixels.  The contours encompass regions covered greater than 12.8, 6.4 and 3.2 seconds of integration per square degree.  The white dots represent the 125 samples taken along an elliptical path for each spin of the flat mirror.  Spins at each end of the azimuth scan are shown along with the dashed boundary of the region observed during one scan of the telescope.  }}
\bigskip
}

\subsection{Flight Performance\label{s_flperf}}

During flight, the HACME telescope performed as expected with no known difficulties.  Once scanning a region, the telescope collects data very efficiently.  For the $\gamma$UMi observation, after removal of $16\%$ of the data due to calibrations, we are able to use $99\%$ of the remainder.  One of the advantages of a balloon environment is that essentially all of the data is useable as there is no ``bad weather".  The power spectra of noise for the three subbands have high frequency (white noise) limits of 0.9, 0.5 and 0.4~mK$\sqrt{\mathrm{sec}}$ with a typical knee frequency of 60~Hz (the frequency at which the noise power is double the white noise limit).  The subbands have typical correlations between 0.5 and 0.7 due to HEMT amplifier gain fluctuations.  This performance is consistent with our laboratory measurements of this receiver and is typical of cryogenic HEMT amplifiers.

\subsection{Systematics}

\subsubsection{Synchronous Offset}

In order to make maps from the data, we need to remove systematic offsets.   When the data are binned into ``sectors" rather than sky pixels, we find an anomolous spin-synchronous offset for each subband.  The peak-to-peak amplitudes of the offsets are 1.3, 1.6 and 4.0~mK in the 39, 41 and 43~GHz subbands, respectively.  The amplitude of these signals varies $<5\%$/hr over the entire flight.  Among the possible causes are atmospheric emission, thermal emission from the optics, the CMB dipole and far sidelobe contamination.  Because the observed offset varies slowly, we are able to remove it on short timescales.  We discuss the possible contributions to this offset and its removal below.  

{
\refstepcounter{myfigs}
\label{f_atmo}
	\centerline{{\vbox{\epsfxsize=9.5cm\epsfbox{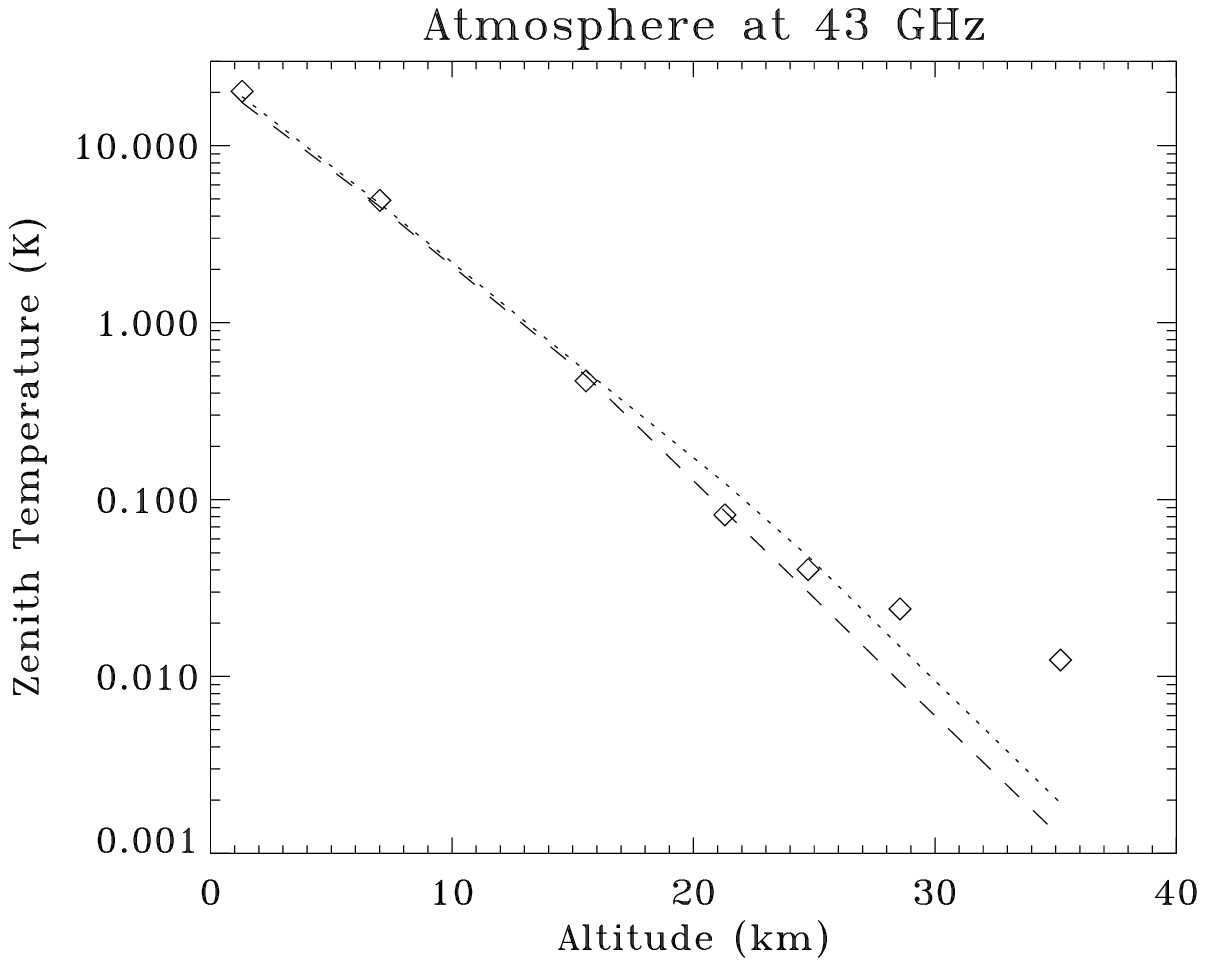}}}}
{\footnotesize{FIG.~\ref{f_atmo}.  Spin-sychronous offset in 43~GHz subband vs. Altitude.  The diamonds are the zenith temperatures assuming the detected offset is due only to the atmosphere.  The dotted line is an atmospheric model prediction using Zhevakin-Naumov line shapes.  The dashed lines is for Van Vleck-Weisskopf lines shapes.  The observation and models differ at float altitude because the atmosphere is no longer the dominant contribution to the offset.}}
\bigskip
}

While the atmosphere is the dominant cause of a spin-synchronous offset on the ground, we find that it is small at our float altitude and remarkably consistent with our predictions at lower altitudes based on a model for atmospheric emission.  We verify the spectrum of the atmospheric contribution and its elevation and altitude dependence up to 25~km with data taken during the ascent of the balloon payload (Figure \ref{f_atmo}).  At an altitude of 35~km, the largest peak-to-peak signals predicted are 500, 600 and 800 $\mu$K for our detection subbands.  During the $\alpha$Leo measurement, the telescope tracked a $22^\circ$ range in elevation, over which the atmospheric signal would have doubled.  With no detectable monotonic change in the offset, the atmospheric signal in the 43~GHz band is limited to less than 600~$\mu$K for this observation.  

Thermal emission from the optics is expected to be a larger contribution to the spin-synchronous offset and opposite in phase from the atmosphere.  Our Q-band receiver detects only one linear polarization and a synchronous signal is expected as the the angle of the flat mirror relative to the receiver changes (\cite{Wol97}).  For our horizontal E-plane, the signal is largest when the beam elevation is greatest.  This signal depends upon the temperature and emissivity of the flat mirror.  The peak-to-peak amplitude of this signal is predicted
to be 7.4, 7.6 and 7.8~mK in the increasing frequency subbands.  The polarization signal is calculated to decrease just $10\%$ over the course of the entire flight as the flat mirror cools.  

We calculate the signal expected from the motion-induced doppler dipole anisotropy (\cite{Line96}).  For a $10^\circ$ throw, the maximum possible signal from the dipole measured by HACME would be $\rm 580 \mu K$.  The synchronous offset produced by the dipole is equal in the three subbands and its phase and magnitude varies with the equatorial coodinates of the target.  

A linear combination of the sources above cannot explain the observed offsets throughout the flight however, so other effects must be present.  Far sidelobe contamination, Radio Frequency Interference, chopped spillover and electrical interference are possibilities.  Because the Earth makes the largest contribution to far sidelobe response, we expect this systematic error to be elevation dependent.  Because the spin-synchronous offset did not change appreciably during the $\alpha$Leo observation, sidelobe contamination is not likely a major cause of systematic error.  Radio Frequency Interference was of great concern to us, due to the necessity of the NSBF flight transmitters and our previous experience.  We were however able to turn off and on all on board transmitters (we even obtained several minutes with {\sl all} transmitters off), and saw no effect on the data.  We cannot entirely rule out chopped spillover and electrical pickup.  Chopped spillover refers to near field off-axis telescope response modulated by the flat mirror or its balance arm.  We performed careful noise integration tests prior to flight to levels far below the observed offset, but differences between ground tests and flight could lead to systematic errors from electrical interference.

\subsubsection{Offset Removal\label{s_offrmv}}

Current data does not allow the separation and identification of the modeled contributions partly due to the cancellation of the effects.  A conservative approach to offset removal is taken because the offsets detected are due to a superposition of the contributions listed above.  We remove all data at the first and second harmonic of the spin rate before analysis, as the offset is expected and observed to be significant only at these harmonics.  Only 3.2\% of the 2~to~3 million degrees of freedom in an observation are sacrificed to this harmonic removal.  We find it a robust approach that sacrifices sensitivity to multipoles $\ell<20$ and use it for all following analysis.

\subsection{Maps}

The total power data collected by HACME are easily converted into sky maps.  Measurements of each ``sector" are assigned a sky position using Eqns.~\ref{thetar} and \ref{phir} and the azimuth and elevation of the center of each flat mirror spin.  Averaging measurements of sky pixels results in maps dominated by stripes due to $1/f$ noise and by the synchronous offset.  The map-making process in \cite{Teg97A} and \cite{Wri96} is needed to reduce striping and excess noise due to the amplifier gain fluctuations and is convenient for the offset removal.  This algorithm accounts for the correlations introduced by $1/f$ noise and optimally reduces striping in the maps by using the inter-connections of our scan pattern.  The same pointing information is used along with $1/f$ noise estimates to make improved maps.  Wiener filtering using signal-to-noise eigenmodes can be used to project out the map modes with the best sensitivity.

Application of this method to these data is described in \cite{Teg99}.  Figure~\ref{f_map} shows a Wiener-filtered map from T99 of the 630 square degree, 3.5 hr, $\gamma$UMi observation, with integration time contours and the scan pattern overlayed.  All structure seen in the map is consistent with instrument noise with pixel errors a factor of 2.6 greater than those expected for the white noise levels indicated above.  The sensitivity is best in the inner, more heavily sampled regions.  T99 find a flat band power upper limit of $\langle \ell(\ell+1)C_\ell/(2\pi)\rangle^{0.5}_B < 77~\mu$K (95\% confidence) at $\ell = 38^{+25}_{-20}$ from this observation.  

\section{Conclusion}
We have built, tested and flown a novel experiment to make large sky area two dimensional maps of the CMB.  The new equipment used at balloon altitudes, the HEMT amplifier and canted spinning flat mirror, worked flawlessly.  We find that atmospheric emission should not significantly contaminate data of future experiments.  The maps produced yield an upper limit to CMB anisotropy because of the poorer sensitivity than expected.  They show no systematic effects or striping from the 1/f noise and demonstrate that the scan stategy used will be an excellent way to make more CMB anisotropy maps.  

\acknowledgments
This work was supported by NASA grant NAGW-1062, CalSpace
Grant CS-55-95, the National Science Foundation, the Center for Particle Astrophysics and NASA grant NAG5-6034.  JS was supported by GSRP grant NGT-51381.  TV, CAW and NF were supported by CNPq grant 910158/95-8.  NF was also partially supported by CAPES.  We would like to thank Geoff Cooke, Glen Schiffrel, and Shea Lovan, the entire crew of the National Scientific Balloon Facility for their effort and dedication, the UCSB Machine Shop for fabricating the newest hardware and Dick Bond, Lloyd Knox, Paolo Natoli and Ned Wright for useful conversations and suggestions.  


\end{document}